# Guidelines for the estimation and reporting of plasmid conjugation rates

Olivia Kosterlitz[1*] and Jana S. Huisman[2,3*]

[1] Biology Department, University of Washington, 3747 West Stevens Way NE, 98195 Seattle, United States of America
[2] Institute of Integrative Biology, Department of Environmental Systems Science, ETH Zurich, Universitätsstrasse 16, 8092 Zürich, Switzerland
[3] Physics of Living Systems, Department of Physics, Massachusetts Institute of Technology, 77 Massachusetts Ave Bldg. 4-332, Cambridge, MA 02139, United States of America (present address)
* To whom correspondence should be addressed: livkost@uw.edu; jhuisman@mit.edu

## Abstract

Conjugation is a central characteristic of plasmid biology and an important mechanism of horizontal gene transfer in bacteria. However, there is little consensus on how to accurately estimate and report plasmid conjugation rates, in part due to the wide range of available methods. Given the similarity between approaches, we propose general reporting guidelines for plasmid conjugation experiments. These constitute best practices based on recent literature about plasmid conjugation and methods to measure conjugation rates. In addition to the general guidelines, we discuss common theoretical assumptions underlying existing methods to estimate conjugation rates and provide recommendations on how to avoid violating these assumptions. We hope this will aid the implementation and evaluation of conjugation rate measurements, and initiate a broader discussion regarding the practice of quantifying plasmid conjugation rates.

# 1. Introduction

Conjugation – the transfer of a plasmid between neighboring bacteria – plays a central role in bacterial ecology and evolution. However, there is little consensus on how to determine plasmid conjugation rates[1,2]. Indeed, the experimental assays and analytical methods commonly used to quantify conjugation differ widely and are often biased[1,3]. In the past year, we have developed new methods to extend the range of biological systems for which conjugation rates can be estimated accurately[1,3]. However, across all available methods, the accuracy and precision of conjugation rate estimates strongly depend on the way experiments are designed and implemented. Clear and detailed reporting is essential to assess the quality of published conjugation rate estimates. Unfortunately, many publications lack the necessary detail to evaluate and reproduce the reported experiments. Here, we propose actionable guidelines for experimental design and reporting, and aim to start a broader discussion about best practices when estimating plasmid conjugation rates.

One important source of confusion is the common use of methods that describe conjugation but do not quantify conjugation rates. These methods (hereafter "population ratios") calculate the ratio between two populations involved in conjugation (e.g., the number of transconjugants per donor). Rather than conjugation alone, this composite metric combines plasmid spread through conjugation and clonal growth into a single quantity. As a result, these measures are affected by environmental conditions that influence growth (e.g., nutrient concentrations or antibiotics). They also vary depending on the initial population densities, initial donor-to-recipient ratio, and duration of the conjugation assay[1,4].

Instead, methods to estimate conjugation rates (e.g., the Levin *et al.* method[5], the Simonsen *et al.* end-point method[4], the Huisman *et al.* approximate Simonsen method[1], or the Kosterlitz *et al.* Luria-Delbrück method[1,3,4]) use explicit models of bacterial population dynamics to derive a measure for conjugation, independent of growth. Thus, in contrast to population ratios, these methods are robust to changes in initial population densities and initial donor-to-recipient ratios. However, all existing methods are bound by their underlying assumptions about the dynamics of conjugation[1,4]. An overview of these theoretical assumptions can be found in supplementary section 1. When assumptions are violated a method may become inaccurate[1,4]. This restricts the range of biological systems that can be explored with a given method and further underscores the importance of reporting all relevant experimental and environmental parameters to reproduce an experiment.

We developed a checklist to guide the planning of laboratory experiments and the reporting of plasmid conjugation rates. We found that methods to estimate conjugation rates exhibit sufficient similarity to curate such a general checklist. The goal is to help researchers design experiments that produce accurate conjugation rate estimates. In addition, the checklist can help other researchers assess the data and the conclusions drawn from it. Guidelines in other fields have helped standardize experimental design and reporting practices, sparked new method development, and increased research reproducibility[6]. Standardized reporting of conjugation experiments will improve the ability to compare conjugation rates estimated for different plasmids, bacteria, and environments. Such a standardized approach may help resolve contradictory claims in the literature (e.g., the effects of antibiotics on the conjugation rate[7–9]) and translate estimates from *in vitro* settings to more complex environments (e.g., the animal gut[10,11]).

We first clarify nomenclature to establish terminology for the checklist, and to promote standardization in the literature. Then, we give the general outline and motivation for the checklist, focusing on the four main parts: experimental design, verification of assumptions, data analysis, and conclusions. Lastly, we discuss some considerations that may extend beyond the current framework of plasmid conjugation experiments. We hope this piece will prompt a broader discussion regarding best practices for estimating plasmid conjugation rates.

## 2. Nomenclature

To aid the standardized reporting of plasmid conjugation experiments, it is important to define a few key terms and steps in this process. *Conjugation* is a process in which a plasmid is transmitted via close contact between a *donor* (D) and *recipient* (R) bacterium. These populations may be different strains or species. Conjugation turns the recipient into a *transconjugant* (T). The transconjugant population generally contains the same chromosome and resident plasmids as the recipient population, as well as the plasmid(s) that are transferred by conjugation from the donor.

The *conjugation rate* (also called *plasmid transfer rate*) describes the number of conjugation events per donor density, per unit time. It parametrizes the horizontal infective spread of a focal plasmid and the corresponding increase in transconjugants. Rates are typically reported as population-level averages. Importantly, the conjugation rate is independent of the increase in transconjugants due to clonal growth. This is in stark contrast to the ratio-based methods that attempt to describe conjugation proficiency using dimensionless ratios between the different populations involved in conjugation (e.g., T/D or T/R). We propose to call these *population ratios*, as their common name "conjugation frequency" could be easily confused with the term "conjugation rate." Population ratios do not quantify conjugation dynamics, but describe relative success of a plasmid in a new host (i.e., T in the numerator is compared to various populations in the denominator). The conjugation rate is specific to the conjugating donor and recipient populations, the focal plasmid, and the environmental conditions[2,10]. Hence a conjugation rate is meaningless without referencing the experimental context and the biological entities involved.

Conjugation rates are measured in the laboratory by performing a *conjugation* (or mating) *experiment*. This typically consists of two parts: (i) an *experimental assay* and (ii) a subsequent *quantification method* used to estimate a conjugation rate from the experimentally measured parameters. The experimental assay typically includes separating the donor, recipient, and transconjugant populations through selection to quantify their population densities (e.g., dilution plating with different antibiotics). The *specificity* of such a selection assay describes its ability to select only the intended cell types, thus decreasing the number of false positives. The *sensitivity* of a selection assay describes its ability to accurately enumerate the intended cell types, thus decreasing the number of false negatives.

Each quantification method requires specific parameters to be measured and thus influences the design of the experimental assay. Although many components of a mating experiment may vary between studies (i.e., quantification method, focal hosts and plasmids, selective conditions), there is sufficient similarity between the various quantification methods that a general checklist can help guide the design and reporting of different conjugation experiments.

## 3. Reporting checklist

The reporting checklist (Table 1) is intended to help authors plan and report experiments to estimate conjugation rates. The construction of the checklist was guided by current literature, which suggests that certain biological variables and experimental steps affect the accuracy and precision of conjugation rate estimates. The items in the checklist are organized into four sections: (i) experimental design, (ii) verification of assumptions, (iii) data analysis, and (iv) conclusions. Here we briefly summarize the motivation and rationale for these general sections of the checklist. A more detailed description and an example of each item are provided in the supplementary material (section 2).

The first section, experimental design, deals with the study's blueprint for estimating conjugation rates. The six items in this section are motivated by the extensive literature showing that conjugation rates depend on (a)biotic factors such as the identity of donors, recipients, and their plasmids[2,10,12,13], the physiology of the donors and recipients[14], spatial structure[15], and temperature[16]. For instance, the conjugation rate often changes depending on the growth phase of the participating donor and recipient bacteria in a plasmid-dependent manner[14]. Therefore, item 1f calls for reporting details on the preparation of donor and recipient bacteria for the mating assay. Additionally, item 1g focuses on the known dependence of conjugation rates on the probability of donor and recipient bacteria to encounter each other. Therefore, the shape and size of the vessels used for the conjugation assay and the shaking speed could affect estimated conjugation rates and should be reported[17]. These details are not only crucial for the replication of the study, but ultimately affect the interpretation of the estimated conjugation rates by authors, reviewers, and readers.

The second section focuses on verifying the experimental and theoretical assumptions that underlie the experimental assay and its quantification. Violations of either type of assumption can lead to inaccurate conjugation rate estimates regardless of the chosen quantification method[1,3,4,15]. Notably, the experimental design (section 1) will determine what assumptions to verify. For instance, a common experimental assumption is that the selective conditions quantify only transconjugants that arose in the mating culture. However, in the case of selective plating, it has been shown that donors and recipients often continue to form transconjugants on the transconjugant-selective plates[18–20]. If this occurs, the transconjugant density is overestimated, and the conjugation rate estimate will be inflated. On the theoretical side, a common assumption is that the donors and transconjugants conjugate to recipients at the same rate. However, suppose the donors and recipients are different species. In that case, the donor-to-recipient (i.e., cross-species) conjugation rate will likely differ from the transconjugant-to-recipient (i.e., within-species) rate. When an experimental or theoretical assumption is violated, authors may need to adjust the experimental design to obtain accurate conjugation rate estimates. Modifications can include changing the quantification method, the conjugation protocol (to minimize the effects of violations), or the biological samples. We recommend checking some assumptions during the experiment, while it may be advisable to check others beforehand. To produce reliable and accurate conjugation estimates, allowing a feedback loop between the verification of assumptions and the experimental design is key.

The third section of the checklist deals with data sharing and analysis. Good data management and reporting serve three main purposes: it helps catch mistakes in data entry or analysis scripts, helps determine whether the analysis was appropriate, and aids other researchers when using or replicating the reported results[21–23]. Data analysis choices profoundly affect the final results and should be documented in as much detail as the experimental protocol. Ever more research institutions, journals, and funders require data sharing according to the FAIR principles (Findability, Accessibility, Interoperability, and Reusability)[22]. These dictate best practices in data sharing including the use of standard, machine-readable file formats, and reporting metadata describing data columns and variables[22–24]. It is important to share not only processed data and research results but also the raw data from experiments. This helps preserve the original data, evaluate the reliability of density estimates, understand the data analysis steps, and reuse the data in future studies such as meta-analyses[21,24]. In addition to data, all workflows, methods, and scripts that led to the final results should be reported and deposited in publicly accessible repositories. We encourage readers to study existing guidelines to improve code and data readability[21,25,26].

The last section of the checklist covers the interpretation of results and conclusions drawn from the experiment. All other items in the checklist should be incorporated into this step. Together they determine the generality of the conjugation rate estimates and their suitability to address the motivating question of the study. Careful consideration of items in the checklist may cause the conclusions of the study to be more limited or conditional than intended. If this information is used early it can still help change the experimental design. Thus, it may be useful to envision this item not necessarily as the last step but as an informative step in a feedback loop.

**Table 1:** Overview of the reporting checklist for conjugation experiments. The checklist items are organized into four main parts (1-4) with multiple items per part (e.g., 1.a-1.g). A brief description is provided for all items. A more detailed description and example for each item are provided in Supplemental Section 1.

| 1. | **Experimental design** | |
|---|---|---|
| 1.a | Purpose | The question, goal, hypothesis, and rationale of the experiment. |

| 1.b | Experimental variables | The experimental conditions used. State the different biological (e.g., bacteria or plasmids) and environmental parameters (e.g., temperature, growth media) used in each assay. |
|---|---|---|
| 1.c | Biological samples | The donors and recipients used. Report strain identity (e.g., taxonomy, sequence accessions, source) and characteristics (e.g., antibiotic resistance profiles). |
| 1.d | Quantification method(s) to be used | The conjugation quantification method(s) applied and the corresponding variables measured during the experiment. |
| 1.e | Description of the selective conditions | The chosen quantification method (e.g., dilution plating, flow cytometry, qPCR) and selective agent(s) (e.g., antibiotics, wavelength for fluorescent markers, primers) used. Include the expected results of each selective agent. |
| 1.f | Sample preparation | The preparation of the biological samples for the assays (e.g., freezer conditions, reanimation procedure, enrichment/growth protocol, growth medium, culturing vessel, added selective agents). |
| 1.g | Protocol details | The full conjugation protocol (e.g., preparation of the mating mixtures and the chosen incubation times). Clearly describe each quality control step (e.g., number of biological/technical replicates, equipment calibrations, controls). |
| **2.** | **Verification of assumptions** | |
| 2.a | Verification of experimental assumptions | The assays used to verify experimental assumptions such as the specificity and sensitivity of the selective conditions and the absence of post-assay conjugation in the selective conditions. |
| 2.b | Verification of theoretical assumptions | The assays or implementation procedures used to verify that the conjugation protocol abides by the theoretical assumptions of the chosen quantification method(s). |
| **3.** | **Data analysis** | |
| 3.a | Raw data | Provide the unprocessed raw data for each assay reported in the study with the appropriate metadata that describes the identity of all variables in the dataset. |
| 3.b | Analysis | The analysis steps used to process the data and to calculate the conjugation rates with the chosen quantification method(s). |

| 3.c | Processed data | Provide the processed data for each figure or analysis reported in the study with the appropriate metadata that describes the identity of all variables in the dataset. |
|---|---|---|
| 3.d | Results | Report results of the conjugation assay including variance across technical replicates. |
| **4.** | **Conclusions** | |
| 4. | Conclusions | Interpret the experimental results in light of the assumptions and limitations of the experimental assay and quantification method(s). |

# 4. Discussion

Conjugation rate estimates are important to understand and predict the ecology and evolution of bacterial communities. Here, we revisited the approaches used to estimate plasmid conjugation rates. We found substantial similarities across existing experimental assays and quantification methods, providing an opportunity for synthesis and general recommendations. First, we propose to unify the nomenclature on plasmid conjugation rates to aid communication in the field. Second, we curated a structured list of things that are important when designing and executing conjugation rate estimation experiments. We recommend adopting this as a reporting checklist while documenting and assessing conjugation experiments. The checklist is a general starting point, but we acknowledge that the content reported for each checklist item will differ substantially between studies. Indeed, there are a large number of possible combinations - including quantification methods, hosts, plasmids, and environments - in the experimental design section of the checklist. As a result, there is no one-size-fits-all approach to executing conjugation experiments.

The checklist is not meant to define a linear path of execution for a conjugation experiment. If an experimenter needs to adjust a design component, it will likely require revisiting other items in the checklist. For example, one may shorten the incubation time of the mating assay to avoid violating the assumption of exponential growth at a constant growth rate. Yet, this in turn can increase the variance in conjugation rate estimates, and decrease the ability to detect differences between conjugation rates. These feedback loops can complicate troubleshooting because each design choice is intertwined with and dependent on others. A seemingly simple modification may cause a cascade of adjustments to other elements of the experimental design.
Though our guidelines don't offer system-specific modifications (although see Supplementary Material section 1 for suggested modifications), the checklist can be a structured aid for designing, troubleshooting, executing, and reporting.

A clear focus for future development should be to address the limitations of existing quantification methods (an overview of their assumptions can be found in Supplementary Material section 1).

Although new methods have become available this past year, there are still combinations of biological questions, hosts, plasmids, and environments that no available method can address. This opens several important avenues for future method development. For example, conjugation in a spatially structured environment, including on agar and filters, violates the assumption of well-mixed populations common to all existing methods. As such, no method can accurately assess conjugation rates in the numerous spatially structured environments bacteria inhabit in nature, including highly relevant ones such as biofilms. Novel techniques like single-cell imaging could help develop new models to quantify the effects of spatial mixing and lead to novel ways to estimate bulk conjugation rates in structured environments[27]. Overall, we recommend that the available techniques to estimate plasmid conjugation rates be continually evaluated. We hope this will spark new method development and make existing methodologies more robust and reliable.

# 5. Conclusion

We provided guidelines for estimating and reporting plasmid conjugation rates. These are based on careful consideration of the sources of variability when estimating such rates and constitute what we consider best practices. We hope this will initiate a broader discussion regarding the practice of estimating plasmid conjugation rates and improve both the implementation and evaluation of conjugation measurements. Ongoing method development will extend the number of environments and species in which we can estimate and compare plasmid conjugation rates, and will also feed into this continuing discussion. We encourage researchers to reconsider these guidelines and the checklist as new insights and innovations become available.

# Acknowledgments

We thank Ben Kerr and Eva Top for their feedback and discussion of the manuscript. This work was supported by the Environmental Biology Division from the National Science Foundation (grant number 2142718 supported O.K.).

# Author contributions

OK: Conceptualization; Investigation; Methodology; Writing - original draft; Writing - review & editing
JSH: Conceptualization; Investigation; Methodology; Writing - original draft; Writing - review & editing

# Supplemental material

## Supplemental section 1: Notes on common theoretical and experimental assumptions

<u>Notes on theoretical assumptions</u>
Choosing a quantification method (item 1.d in Table 1) will dictate which theoretical assumptions need to be verified (item 2.b in Table 1). Since each method differs slightly in its underlying assumptions, we provide a brief overview of the modeling assumptions (see Table S1) for several quantification methods including Levin *et al.* (hereafter TDR)[5], Simonsen *et al.* (SIM)[4], Huisman *et al.* (hereafter ASM)[1], and Kosterlitz *et al.* (LDM)[3].

Table S1: Overview of modeling assumptions for several quantification methods. A cross (X) indicates that this is an assumption of the given method.

| **Assumption** | TDR | SIM | ASM | LDM |
|---|---|---|---|---|
| A. Conjugation events follow mass-action kinetics. | X | X | X | X |
| B. The plasmid loss rate of the focal plasmid is zero. | X | X | X | X |
| C. The growth rate is identical for all cell types. | X | X | | |
| D. The number of conjugation events from transconjugants is low relative to the conjugation events from donors. | X | X | X | |
| E. Growth and conjugation respond in a functionally similar way to changes in resources. | | X | | |
| F. The populations do not change in size due to growth. | X | | | |
| G. The populations grow exponentially at a constant growth rate. | | | X | X |

Here, we provide additional details for each modeling assumption in Table S1.

*(A) Conjugation events follow mass-action kinetics* (assumed by TDR, SIM, ASM, and LDM). We recommend running conjugation assays in a liquid medium with shaking. Mating assays cultured in structured environments (i.e., on agar or filters) violate this assumption unless densities on filters are high and the mating duration is short[15]. Since each existing quantification method makes this assumption, the current methods are not appropriate for exploring various effects of environmental structure on conjugation. Structure breaks the homogeneity needed to satisfy this mass-action kinetics assumption. In environments that are less well mixed, the variation across multiple replicate mating assays will typically be higher than in well-mixed environments.

*(B) The plasmid loss rate of the focal plasmid is zero* (assumed by TDR, SIM, ASM, and LDM). Simulations[3,4] have shown that violations to this assumption have minimal effect if the conjugation assays are short (i.e., less than 24 hours) even with high segregation rates. However, some plasmid loss scenarios remain unexplored, such as plasmid incompatibility between the incoming donor plasmid and a resident plasmid.

*(C) The growth rate is identical for all cell types* (assumed by TDR and SIM).
Transfer rate estimates can become biased due to differences in donor, recipient, and transconjugant growth rates[1,3,4]. There are two general suggestions when the growth rate differences cause non-negligible inaccuracies. First, it may be possible to continue using the TDR and SIM methods by shortening the assay time as much as possible. However, other problems - such as high measurement noise - can arise when shortening the assay time. Second, one can choose a method that allows for heterogeneous growth rates such as the ASM or LDM.

*(D) The number of conjugation events from transconjugants is small relative to the number of conjugation events from donors* (assumed by TDR, SIM, and ASM). The transfer rate estimates will become increasingly biased when conjugation events become dominated by the transconjugant population. This will occur more quickly when the transconjugant population grows rapidly, and when the conjugation rate from transconjugants is much higher than that from donors. Violations of this assumption can introduce some of the highest magnitude bias. Again, two solutions exist: (i) one can shorten the assay time, or (ii) use a method that focuses on the phase of conjugation where transconjugant populations are small. The ASM focuses on the first point, by providing system-specific estimates of the critical incubation time when this assumption will start to be violated. The LDM follows the second strategy, as it was specifically designed such that this assumption always holds.

*(E) Growth and conjugation respond in a functionally similar way to changes in resources* (assumed by SIM).
The SIM method was created and implemented based on the assumption that conjugation rate is maximal during exponential growth. Thus it is assumed that the growth rate ramps up (out of lag phase into exponential) and down (out of exponential into stationary phase) proportionally to the conjugation rate. This means the assay can run for 24 or 48 hours and produce a maximum conjugation rate estimate. This assumption is modeled on the behavior of F plasmids[5] but is known to be violated by other plasmid families[14]. When this assumption does not hold, more or less transconjugants will exist at the end of the assay than expected by the model, leading to inaccurate transfer estimates. Confirmation of this assumption requires time course measurements and greatly increases the difficulty of the assay. Therefore, we suggest avoiding this assumption by running the assay over a short period and mixing donors and recipients in exponential growth (assuming the focal plasmid conjugates during the exponential growth phase). These growth conditions (i.e., controlling the bacterial growth phase throughout the experiment) are those required by the ASM and LDM. In other words, we suggest avoiding any assumptions about how growth and conjugation relate to one another and implementing the conjugation assay such that growth is controlled and kept as constant as possible.

*(F) The bacterial populations do not change due to growth* (assumed by TDR).
To ensure negligible growth over the assay period, measure the initial donor and recipient densities to calculate the growth rate during the assay (which should be very close to zero).

*(G) The cell populations grow exponentially at a constant rate* (assumed by ASM and LDM). There are several approaches to ensure a relatively constant growth rate over the assay period. One option is to measure the donor and recipient growth rates during the assay by estimating the donor and recipient densities at various midpoints. Another option is to test the protocol conditions before setting up the conjugation assay.

Notes on experimental assumptions
We note that some strategies to avoid violating the theoretical assumptions of a method may introduce experimental difficulties. Indeed, researchers may find that removing violations of theoretical assumptions can result in violating experimental assumptions. For instance, a common solution to differences in growth and conjugation affecting the conjugation rate estimates is to shorten the assay duration. However, this can result in measurements of very few transconjugants. When using selective plating, very low dilutions must be used to find the few existing transconjugants. This increases the density of donors and recipients that will be plated on transconjugant-selecting agar plates and thus the chance of transconjugants arising due to conjugation between donors and recipients on these agar plates. Shortening the assay duration also increases the variance [3]. This may result in detecting no transconjugants in some replicate mating assays. For most estimates (besides the LDM), a transconjugant density estimate of zero leads to a zero estimate for the conjugation rate. If this occurs in one out of three replicates, this zero data point is providing information about the conjugation rate and when averaged with the other two replicates will improve the accuracy of the estimate. However, a high variance may limit the ability to detect differences in conjugation rates. Therefore, assay modifications have inherent tradeoffs. Balancing these sources of error may be more difficult for certain methods and chosen biological samples.

Besides post-assay conjugation, there are other experimental assumptions (item 2.a in Table 1) that should be verified including the specificity and selectivity of the selective conditions. Specificity assumes that each population's selective conditions enriches only the target population, not the others. For example, the donor-selecting condition should only select donor cells and reliably eliminate recipients and transconjugants. Sensitivity instead assumes that each population's selective condition reliably enriches all the cells of the target population and not only a proportion of these cell types. For example, a common way to enrich particular populations is to use antibiotics. Recently it has been shown that antibiotics don't reliably enrich all resistance cells in a bacterial culture and the extinction of resistant cells can be substantial[28] leading to an underestimate of the target population. Thus, if the antibiotics chosen to enrich donors, recipients, and transconjugants have different extinction probabilities across these populations, the estimates of population densities will be inaccurate. This will produce a systematic error when calculating the conjugation rate.

We generally recommend to verify all theoretical and experimental assumptions to ensure the estimated conjugation rates are accurate.

## Supplemental section 2: Detailed description and examples of the reporting checklist

The purpose of the reporting checklist is to help researchers provide enough detail for independent validation and interpretation of their experiments. Detailed descriptions and examples (in blue text) for the reporting checklist are provided below. The examples are adapted from the conjugation experiments reported in Kosterlitz *et al.*[3] For brevity, we do not include all of the details from the study. For example, Kosterlitz *et al.* reported conjugation experiments using both the Luria-Delbrück method[3] (hereafter LDM) and the Simonsen *et al.* method[4] (hereafter SIM). However, the examples for some items only focus on estimating the conjugation rate with the LDM approach and do not include descriptions of the SIM approach and vice versa.

**1)    Experimental Design**
a)    Purpose
A clear description of the experiment's question, goal, hypothesis, and rationale.
The purpose of this experiment was to compare two quantification methods (SIM vs. LDM) to estimate plasmid transfer rates by quantifying the cross-species (*Klebsiella pneumoniae* and *Escherichia coli*) and within-species (*E. coli* to *E.coli*) plasmid transfer rate for an IncF plasmid using the SIM and LDM assays.

b)    Experimental variables
A clear description of the experimental conditions used. Specify the different biological samples (e.g., different bacteria or plasmids) and environmental parameters (e.g., temperature, growth medium). Make explicit which conditions vary between treatments, and which stay fixed (especially when such conditions could vary in other studies).
There are two treatment groups (T1 and T2) which differ with regards to the donor species (*K. pneumoniae* and *E. coli*, respectively). The same recipient (*E. coli*) and focal plasmid (IncF) were used in both treatments. All experiments were carried out at 37 °C in LB medium in 96-well deep-well plates on a Bellco Biotechnology mini-orbital shaker and shaken at 400 rpm.

c)    Biological samples
A clear description of each donor and recipient strain (e.g., taxonomy, sequence accessions, source of the sample) including the strain characteristics (e.g., antibiotic resistance profiles) as they pertain to the experiment design.
Table S2: A description of the donor, recipient, and plasmid used in the T1 treatment. The strain identity is abbreviated (Strain abb. column) using the first letter of the host genus to indicate the host (i.e., E or K) and a shorthand in parenthesis indicating the plasmid state (plasmid-free, Ø, or plasmid-containing, pF). The antibiotic resistance (ABR) profiles are indicated using antibiotic

abbreviations (tet = tetracycline, str = streptomycin, and nal = nalidixic acid) where the 'R' superscript indicates drug resistance in the strain.

| Strain abb. | Host | Plasmid | ABR profile | Detailed description (i.e., source, sequence accessions) |
|---|---|---|---|---|
| K(pF) | *K. pneumoniae* Kp08 nal[R] | F'42 tet[R] | nal[R], tet[R] | The host[18] and plasmid[29] are described in previous studies. |
| E(Ø) | *E. coli* K-12 BW25113 str[R] | None | str[R] | *E. coli* K-12 BW25113 was marked with streptomycin resistance using selective plating in a previous study[30]. |

d)  Quantification method(s) to be used

Clearly state which quantification method(s) will be applied, define the variables to be measured during the experiment (e.g., $D_0$, $D_{\bar{t}}$, etc.), and provide a verbal description of each variable.

For this experiment, we used two methods: LDM and SIM. A description of the measured variables is given in Table S3.

Table S3: Variables to be measured in the laboratory.

| Required measurements | Relevant method | Description |
|---|---|---|
| $\bar{t}$ | SIM, LDM | The incubation time at which the conjugation assay is terminated, and the final variables are measured. |
| $D_0$, $D_{\bar{t}}$ | SIM, LDM | Initial and final donor density, respectively. |
| $R_0$, $R_{\bar{t}}$ | SIM, LDM | Initial and final recipient density, respectively. |
| $T_{\bar{t}}$ | SIM | Final transconjugant density. |
| $\hat{p}_0(\bar{t})$ | LDM | Fraction of replicate donor-recipient cocultures without transconjugants at the end of the assay. |

e)  Description of the selective conditions for donors, recipients, and transconjugants.

Clearly state the chosen selection method (e.g., dilution plating, flow cytometry, qPCR) and selective agent(s) (e.g., antibiotics, wavelength for fluorescent markers, primers) to separate different bacterial populations. This should include the details necessary to reproduce and interpret the experimental results. Specifically, the description of the selection scheme should include each selective agent's expected results and how these will produce the required measurements (e.g., $D_0$) to calculate the conjugation rate using the chosen method(s).

We used dilution plating supplemented with selective antibiotics (see S4) to estimate the initial ($t = 0$) and final ($t = \bar{t}$) densities for the donors ($D_0$ and $D_{\bar{t}}$), recipients ($R_0$ and $R_{\bar{t}}$), and transconjugants ($T_{\bar{t}}$, we note $T_0$ = 0). To estimate the initial and final densities of the total population ($N_0$ and $N_{\bar{t}}$), we took the sum of all sub-populations. To determine the fraction of replicate mating populations that have no transconjugants at the final incubation time ($\hat{p}_0(\bar{t})$), we added liquid medium supplemented with selective antibiotics to replicate donor-recipient cocultures in a deep-well plate and screened for turbidity after an additional incubation period.

Table S4: Expected outcomes in the chosen selective conditions. Antibiotic abbreviations follow the convention used in Table S3. The chosen selective conditions permit the growth of transconjugants in the selective conditions used to estimate the densities of donors and recipients. Therefore, we correct the donor and recipient density estimates by subtracting the transconjugant density estimated from the transconjugant-selecting agar plates. Notably, this correction does not make a meaningful difference in our case since the final transconjugant density is orders of magnitude lower than that of donors and recipients.

| Method | Required measurement(s) | Description of selection | | | Expected selective outcomes | | |
|---|---|---|---|---|---|---|---|
| | | Selective agent | Purpose | Media type | Donor K(pF) | Recipient E(Ø) | Transconjugant E$_T$(pF) |
| SIM, LDM | $D_0, D_{\bar{t}}$ | tet | Selection for focal plasmid | Agar | Colony formation | No colony formation | Colony formation |
| SIM, LDM | $R_0, R_{\bar{t}}$ | str | Selection for recipient host | Agar | No colony formation | Colony formation | Colony formation |
| SIM | $T_{\bar{t}}$ | tet + str | Selection for focal plasmid and recipient host | Agar | No colony formation | No colony formation | Colony formation |
| LDM | $\hat{p}_0(\bar{t})$ | tet + str | Selection for focal plasmid and recipient host | Liquid | Non-turbid culture | Non-turbid culture | Turbid culture |

f) Sample preparation

Clearly describe how the samples were prepared for the assays, including the freezer conditions, reanimation procedure, enrichment/growth protocol, growth medium, culturing vessel, and the added selective agent(s) if any.

The strains were inoculated into 250µl LB medium from frozen glycerol stocks and grown overnight. The plasmid-containing cultures were supplemented with 15 µg ml$^{-1}$ tetracycline to maintain the plasmid. The saturated cultures were diluted 100-fold into LB medium to initiate a second 24 hours of growth (to acclimate the previously frozen strains to laboratory conditions). This procedure was used to prepare all strains before each assay.

g) Protocol details

Clearly describe the conjugation protocol. This should include the details necessary to replicate the experiments such as the preparation of the mating mixtures and the chosen incubation time ($\tilde{t}$). Clearly describe the growth medium, culturing vessel, and the added selective agent(s) if any. In addition, clearly describe each quality control step such as the number of biological/technical replicates, equipment calibrations, controls within the assay, etc.

The prepared strains (described in item 1.f) were diluted 10,000-fold into LB medium and incubated for 4 hours to ensure the cultures entered exponential growth. The exponentially growing cultures were diluted 1,000-fold and mixed at equal volumes to create a large volume of donor-recipient coculture. The coculture was dispensed into 84 wells of a deep-well microtiter plate at 100 µl per well to create the replicate cocultures used to calculate $\hat{p}_0(\tilde{t})$. Three wells (W1-3) of the same deep-well microtiter plate were used to determine the initial densities ($D_0$ and $R_0$) via selective plating: 130 µl of the coculture was dispensed per well, and immediately 30 µl was removed for plating. Three additional wells were used to determine whether the transconjugant-selecting medium prohibited the growth of donors and recipients while permitting the growth of transconjugants. These three wells contained 100 µl of donor, recipient, and transconjugant cultures for growth in monoculture (MW1-3). Lastly, four wells of the deep-well microtiter plate contained additional diluted monocultures of donors and recipients (2 wells each at 100 µl). These monocultures were used to create fresh cocultures (in the two remaining empty wells, EW1-2) at the final incubation time (see below). The deep-well plate was incubated until the final incubation time ($\tilde{t}$ = 4h), after which three events occurred in rapid succession. First, 30 µl was removed from each of the three wells used to determine initial population densities (W1-3) to determine the final population densities ($D_{\tilde{t}}$ and $R_{\tilde{t}}$) via selective plating. Second, into the two empty wells (EW1-2), 50 µl from one replicate of the donor and recipient monoculture were mixed at equal volumes to create a fresh coculture. Later in the assay, non-turbidity in these two wells would verify that new transconjugants did not form via conjugation after adding the transconjugant-selecting medium. Third, 900 µl of transconjugant-selecting medium (with a final concentration of 7.5 µg ml$^{-1}$ tetracycline and 25 µg ml$^{-1}$ streptomycin) was added to all relevant wells (the 84 cocultures used to estimate $\hat{p}_0(\tilde{t})$, the monoculture controls MW1-3, and the freshly mixed coculture controls EW1-2). The deep-well plate was incubated for four days, and the state of all wells (turbid or non-turbid) was recorded at the end. For the selective plating, the agar plates were incubated for 24 hours, colonies were counted, and the counts were recorded. The conjugation protocols were repeated six times on different days.

**2) Verification of assumptions**

a) Verification of experimental assumptions

Clearly describe the assays used to verify that the chosen selective conditions are specific and sensitive to the bacterial populations studied. In addition, describe the assays used to verify that conjugation does not continue to occur after the mating experiment and during the selection step (hereafter "post-assay conjugation"). This should include the necessary details to replicate these assays. The chosen selective conditions should allow accurate measurement of the variables needed to calculate the conjugation rate with the chosen quantification method(s) (see item 1.d). See supplementary section 1 for more notes on experimental assumptions.

<span style="color:blue">The specificity and sensitivity of the chosen antibiotics to select the relevant bacterial populations (e.g., donors) were validated for each antibiotic in the relevant media type (i.e., agar versus liquid). To verify the absence of post-assay conjugation, we created fresh cocultures of donors and recipients at the end of the mating assays to verify that no transconjugant cells were formed after antibiotic exposure (see item 1.g for more details).

We used standard microbiology techniques to verify the specificity of antibiotic selection with agar plates. Specifically, previously prepared strains (see item 1.f) of each population were diluted and plated onto each type of selective plate (donor, recipient, and transconjugant) at the appropriate concentrations (7.5 µg ml$^{-1}$ tet, 25 µg ml$^{-1}$ str, and 7.5 µg ml$^{-1}$ tet + 25 µg ml$^{-1}$ str, respectively). The agar plates were incubated for 24 hours, colonies were counted, and the counts were recorded. Colony formation followed the expected selective outcomes (see S4). To determine the specificity of antibiotic selection in the liquid medium, we conducted a standard minimum inhibitory concentration (MIC) assay to determine the lowest concentration that would inhibit growth of donors and recipients while permitting the growth of transconjugants. Briefly, prepared strains for each population were diluted 100-fold and inoculated in a deep-well plate containing a gradient of transconjugant-selecting medium (a series of 2-fold dilutions of the transconjugant-selecting antibiotics). The deep-well plate was incubated for 24 hours, and the state of the wells (turbid or non-turbid) was recorded. The minimum inhibitory concentration was determined for each population (Table S5), and a concentration specific to the growth of the transconjugant population was chosen (7.5 µg ml$^{-1}$ tet + 25 µg ml$^{-1}$ str).

For the sensitivity protocols, we adjusted the protocol by Alexander and MacLean[28] to estimate extinction probabilities in the antibiotic concentrations identified with the specificity protocols described above. Specifically, to verify the sensitivity of selection on agar plates, prepared strains for each population were diluted and plated onto an antibiotic-free plate and the appropriate selective plate. This was repeated in triplicate for each population. The agar plates were incubated for 24 hours, colonies were counted, and the counts were recorded. To verify the sensitivity of transconjugant selection in liquid medium, the prepared transconjugant strain was diluted 4 x 10$^7$-fold and inoculated into all 96 wells of a deep-well plate containing antibiotic-free medium and a deep-well plate containing the transconjugant-selecting medium (final concentration of 7.5 µg ml$^{-1}$ tet + 25 µg ml$^{-1}$ str) to a final volume of 1 ml. The deep-well plates were incubated for 4 days, and the state of the wells (turbid or non-turbid) was recorded.</span>

Table S5: The dual-drug gradient MIC for each population in both treatments. The MIC was the lowest concentration that produced a non-turbid culture.

| Population type | MIC with a str and tet gradient |
|---|---|
| Donor K(pF) | 1.88 µg ml$^{-1}$ tet + 6.25 µg ml$^{-1}$ str |
| Recipient E(Ø) | 3.75 µg ml$^{-1}$ tet + 12.5 µg ml$^{-1}$ str |
| Transconjugant $E_T$(pF) | 15 µg ml$^{-1}$ tet + 50 µg ml$^{-1}$ str |

b)  Verification of theoretical assumptions

Clearly describe the assays or implementation procedures used to verify that the conjugation protocol abides by the theoretical assumptions of the chosen quantification method(s). If laboratory assays were conducted to validate a theoretical assumption, then the reporting should include the necessary details to replicate these assays, the raw data, and a description of the data analysis. If an implementation procedure in the conjugation protocol is used to conform to a theoretical assumption, the authors should clearly state how the implementation abides by the relevant assumption. In general, the authors should state that their conjugation protocol can produce accurate laboratory estimates for the conjugation rate given that the theoretical assumptions of the chosen conjugation method(s) were met. See supplementary section 1 for more notes on theoretical assumptions.

The LDM makes the following assumptions: (i) the dynamics of conjugation follow mass action kinetics, (ii) the focal plasmid is not lost through segregation, and (iii) the bacterial populations grow exponentially at a constant growth rate. To ensure assumption (i) was met, we ran the conjugation assays in 100 µl liquid medium in 96-well deep-well plates on a Bellco Biotechnology mini-orbital shaker and shaken at 400 rpm. For assumption (ii), simulations have shown that plasmid loss through segregation is negligible during the incubation times used in this study even for high rates of segregational loss[3,4]. Thus, we concluded that violating this assumption would have a negligible effect. To ensure assumption (iii) was met, we verified that a 4-hour pre-assay growth period for both donors and recipients caused cultures to enter exponential growth before initiating the mating assays. Second, we verified that the growth rate of the donors and recipients during the mating assay was similar to the exponential growth rate after the 4-hour growth period by calculating the donor and recipient growth rates over the assay period using the estimated initial and final population densities. Thus, we concluded that the donor and recipient growth rate were relatively constant over the mating assay period.

3)  **Data analysis**

a)  Raw data

Provide the raw data for each assay reported in the study. This is the data that was collected before being processed, cleaned, or analyzed. It is recommended to store this data separately from the file used for analysis, to avoid accidentally changing the original data[21]. The raw data should be accompanied by metadata that describes the identity of all variables in the dataset.

The raw data is provided for each assay in this study (see Table S6; note, this would typically be supplied as a supplementary csv file). A description of each column can be found in the accompanying README file.

Table S6: The header of a csv file containing the raw data from the mating assays including a row containing column names (see the README.file S6 for description) and several rows with data entry.

| Date | Experiment | Treatment | Culture | Time | Replicate | Tet | Str | Population | Dilution | Counts |
|---|---|---|---|---|---|---|---|---|---|---|
| 2020-11-09 | SIM_Kp | T1 | Mating | 0 | 1 | 7.5 | 0 | D | 1 | 18 |
| 2020-11-09 | SIM_Kp | T1 | Mating | 0 | 1 | 0 | 25 | R | 1 | 22 |
| 2020-11-09 | SIM_Kp | T1 | Mating | 0 | 1 | 7.5 | 25 | T | 0 | 0 |
| 2020-11-09 | SIM_Kp | T1 | Mating | 0 | 2 | 7.5 | 0 | D | 1 | 27 |
| 2020-11-09 | SIM_Kp | T1 | Mating | 0 | 2 | 0 | 25 | R | 1 | 19 |
| … | … | … | … | … | … | … | … | … | … | … |

README.file S6:
Date: date the experiment was performed, in YYYY-MM-DD format.
Experiment: a descriptive name for the experiment performed.
Treatment: the treatment identifier. The two treatments, T1 or T2, differ in the identity of the donor species (*K. pneumoniae* or *E. coli*, respectively).
Culture: the type of bacterial culture inoculated onto the agar plate. Takes the values "Mating" for the mating culture, "Donor" for the donor monoculture, or "Recipient" for the recipient monoculture.
Time: the time at which selective plating was performed. Takes two values, either $t = 0$ or $t = \bar{t}$.
Replicate: the identifying number for each technical replicate. Biological replicates were performed on separate days.
Tet: the concentration of tetracycline antibiotic in ug ml$^{-1}$ that was added to the agar plate.
Str: the concentration of streptomycin antibiotic in ug ml$^{-1}$ that was added to the agar plate.
Population: this refers to the donor (D), recipient (R), or transconjugant (T) population to be quantified on the agar plate.
Dilution: the ten-fold dilution factor used on the bacterial culture before inoculating it onto the agar plate.
Counts: the number of colony-forming units counted on the agar plate after incubation.

b)    Analysis

Clearly describe each step used to analyze the data and calculate the conjugation rate with the chosen method(s). This should include the details for data cleaning, processing steps, and statistics. Original scripts or Excel macros used for analysis should be published with the manuscript. Specify the exact formula or software package (including the version number) used to estimate conjugation rates. If existing software packages or tools were used, cite accordingly.

Before calculating the SIM estimate for the conjugation rate, we analyzed the data from the assays in 2.a. Specifically, for the SIM estimate, we need to determine the extinction probabilities of each population on the relevant agar selective condition. The agar extinction probability ($\pi_{x,agar}$) was calculated as:

$$\pi_{x,agar} = 1 - \frac{C_X}{C_0},$$

where $C_X$ is the number of colonies on the antibiotic plate and $C_0$ is the number of colonies on the antibiotic-free plate for the same diluted culture. The extinction probabilities were calculated for each replicate and then averaged. Given the non-zero extinction probabilities on the selective-agar plates (Table S7), we used each cell type's estimated average extinction probabilities to correct the corresponding population's density estimates from the mating assay.

We used several sequential analysis steps to calculate the SIM estimate for the conjugation rate of any donor-recipient dyad. First, using the relevant plating information (see item 3.a) including plate type, ten-fold dilution factor, and colony counts, we calculated the population densities for each technical replicate. Using the estimated extinction probabilities (Table S7), we corrected the density estimates. Next, we averaged the density estimates across the technical replicates for each biological replicate. Next, we calculated the donor-to-recipient conjugation rate for each biological replicate using the SIM equation:

$$\gamma = \frac{1}{t}\left[\ln\left(1 + \frac{T_t}{R_t}\frac{N_t}{D_t}\right)\right]\frac{\ln N_t - \ln N_0}{N_t - N_0}$$

by using the incubation time ($t$), the initial and final density of all bacteria ($N_0$ and $N_t$, respectively), and the final density of each population ($D_t$, $R_t$, and $T_t$). Lastly, we averaged the SIM estimate from each biological replicate to produce an average SIM estimate for the donor-recipient dyad (i.e., treatment).

Table S7: Population-specific extinction probabilities for selective agar plates. Throughout the study donor-, recipient-, and transconjugant-selective plates were prepared at concentrations: 7.5 µg ml$^{-1}$ tet, 25 µg ml$^{-1}$ str, and 7.5 µg ml$^{-1}$ tet + 25 µg ml$^{-1}$ str, respectively.

| Cell type | Selective-plate type | Extinction probability ($\pi_x$) |
|---|---|---|
| Donor K(pF) | Donor | 0.21 |
| Recipient E(Ø) | Recipient | 0.55 |
| Transconjugant E$_T$(pF) | Transconjugant | 0.99 |

c) Processed data

Provide the processed data for each figure or analysis reported in the study. The processed data should be accompanied by metadata that describes the identity of all variables in the dataset.

The data to recreate all figures in this paper can be found in Table S8 (this will typically be included as one or more csv files per experiment) A description of each column can be found in the accompanying README file.

Table S8: The header of a csv file containing the processed data from the mating assays including a row containing column names (see the README.file S8 for description) and several rows with data entry.

| Treatment | Method | Replicate | Estimate |
|---|---|---|---|
| T1 | SIM | 1 | 3.23e-11 |
| T1 | SIM | 2 | 4.71e-11 |
| T1 | SIM | 3 | 7.96e-11 |
| T1 | SIM | 4 | 4.58e-12 |
| T2 | SIM | 5 | 7.84e-12 |
| … | … | … | … |

README.file S8:

Treatment: the treatment identifier. The two treatments, T1 or T2, differ in the identity of the donor species (*K. pneumoniae* or *E. coli*, respectively).
Method: the quantification method used to estimate the conjugation rate.
Replicate: the identifying number for each biological replicate. Technical replicates were averaged into one value per biological replicate.
Estimate: estimated value of the conjugation rate.

d) Results

Report measurements and corresponding conjugation rate estimates with the variance across replicates. Biological replicates of the mating assay are important since conjugation is a stochastic process. Always depict and report estimates with appropriate error bars, and ideally show replicate estimates of the conjugation rate for a given donor-recipient dyad as individually plotted points. For estimates that are derived from multiple variables that are each individually measured and associated with an error rate ( most formulas to estimate conjugation rates contain measured parameters such as population densities and growth rates), the error should be propagated using Gaussian error propagation.

The LDM estimate for the cross-species conjugation rate ($1.96 \times 10^{-13} \pm 2.06 \times 10^{-13}$) was significantly lower than the SIM estimate ($4.98 \times 10^{-10} \pm 4.53 \times 10^{-10}$) by approximately three orders of magnitude (t-test, $p < 0.0001$). In contrast, the LDM estimate for the within-species conjugation rate (1.42 x $10^{-7}$ ± 1.07 x $10^{-7}$) was statistically insignificant (t-test, p > 0.05) compared to the SIM estimate (9.86 x $10^{-7}$ ± 1.08 x $10^{-6}$).

## 4) Conclusions

A summary of the authors' interpretation of the experimental results. This integrates information from parts 1-3, as the experimental assay and conjugation rate estimation method will come with assumptions and limitations that guide the interpretation of the results.

The SIM estimate for the cross-species conjugation rate is inflated compared to the LDM estimate. Simulations have shown that an elevated within-species conjugation rate (i.e., transconjugant to recipient conjugation rate) violates a theoretical assumption of the SIM approach and can lead to an inflated and inaccurate SIM estimate for the cross-species conjugation rate (i.e., donor-to-recipient conjugation rate). Notably, the accuracy of the LDM approach is not affected by asymmetry between these rates (donor and transconjugant conjugation rates). Indeed, the SIM and LDM estimates for the within-species conjugation rate are consistent with this explanation.